\documentclass[aps,prl,preprint,superscriptaddress]{revtex4-1}
\usepackage{graphicx,amsmath,mathrsfs}

\begin{document}


\title{Towards optical frequency comb generation in continuous-wave pumped titanium indiffused lithium niobate waveguide resonators}


\author{Michael Stefszky}
\affiliation{Integrated Quantum Optics, Applied Physics, Paderborn University, Warburger Strasse 100, 33098 Paderborn, Germany}
\affiliation{Molecular Science, Department of Chemistry, P.O. Box 55 (A.I. Virtasen aukio 1), FI-00014 University of Helsinki, Finland}

\author{Ville Ulvila}
\affiliation{Molecular Science, Department of Chemistry, P.O. Box 55 (A.I. Virtasen aukio 1), FI-00014 University of Helsinki, Finland}

\affiliation{VTT Technical Research Centre of Finland Ltd., Vuorimiehentie 3, 02150 Espoo, Finland}

\author{Zeina Abdallah}
\affiliation{Laboratory of Photonics, Korkeakoulunkatu 3, Tampere University of Technology, FI-33720, Tampere, Finland}

\author{Christine Silberhorn}
\affiliation{Integrated Quantum Optics, Applied Physics, Paderborn University, Warburger Strasse 100, 33098 Paderborn, Germany}

\author{Markku Vainio}
\affiliation{Molecular Science, Department of Chemistry, P.O. Box 55 (A.I. Virtasen aukio 1), FI-00014 University of Helsinki, Finland}
\affiliation{Laboratory of Photonics, Korkeakoulunkatu 3, Tampere University of Technology, FI-33720, Tampere, Finland}


\date{\today}

\begin{abstract}
Much progress, both experimentally and theoretically, has recently been made towards optical frequency comb generation from continuously pumped second-order nonlinear systems. Here, we present observations towards finding an \textit{integrated} solution for such a system, using a titanium indiffused lithium niobate waveguide resonator. These results are compared to recently developed theory for equivalent systems. The system is seen to exhibit strong instabilities, which require further investigation in order to fully determine the suitability of this platform for stable optical frequency comb generation.
\end{abstract}


\maketitle

\section{Introduction}

Optical frequency combs (OFC's) are highly broadband coherent light sources that have found uses in a range of applications such as time and frequency metrology and molecular spectroscopy \cite{Holzwarth00.PRL,Gherman02.OE}. The usefulness of OFC's has led to rapid commercialization of the technology. The most common and reliable method of OFC generation has been through the use of mode-locked lasers, which provide high stability over long time time scales and can span large frequency ranges \cite{Cundiff03.RMP,Diddams99.OL,Fortier06.OL}. However, these systems tend to be quite large, and therefore there has been a strong drive towards miniaturization of the technology.

Recently, most of the work regarding this goal has been directed towards systems involving the third-order nonlinearity. For example, OFC's have been produced in optical fiber cavities \cite{Obrzud17.NP}, whispering gallery mode resonators \cite{Savchenkov08.PRL} and a number of microresonator systems \cite{Xue17.LSA,DelHaye07.N}. These are all promising systems, but they are also often plagued by extraneous effects such as mode competition and photothermal instabilities due to the high intensities required for comb generation (typically achieved through the use of strong pulsed light) \cite{Xue17.LSA}. A deeper understanding of these processes and possible methods for mitigating them are active areas of research \cite{Saha13.OE,Carmon04.OE,Liu14.O,DelHaye08.PRL}.

More recently, OFC's have also been observed from continuous-wave (CW) pumped second-order nonlinear systems. These systems typically offer nonlinear interaction strengths that are orders of magnitude larger than those found in third-order nonlinear systems. This may provide one avenue for reducing the amount of intra-cavity power required for comb generation, thereby providing some method of reducing unwanted thermal effects and thermally-driven mode competition. In addition, CW lasers are often simpler to work with than their pulsed counterpart. OFC generation in such systems typically occurs due to a number of cascaded nonlinearities between the various frequency components; second-harmonic generation (SHG) and optical parametric oscillation (OPO) \cite{Ulvila15.PRA}. The general scheme is shown in Fig. \ref{CombGen}.  CW pumped OFC generation utilising the second-order nonlinearity has been observed in various free-space cavity systems \cite{Ulvila13.OL,Ulvila14.OE,Ulvila15.PRA,Ulvila13.OL,Ricciardi15.PRA,Mosca18.PRL}. The recent results from CW pumped resonator systems utilising the second-order nonlinearity for comb generation has led to more complete theories that can explain much of the behavior seen from these devices \cite{Leo16.PRL,Hansson17.PRA}. 

\textbf{\begin{figure}[!ht]
		\centering
		\includegraphics[width = 10cm]{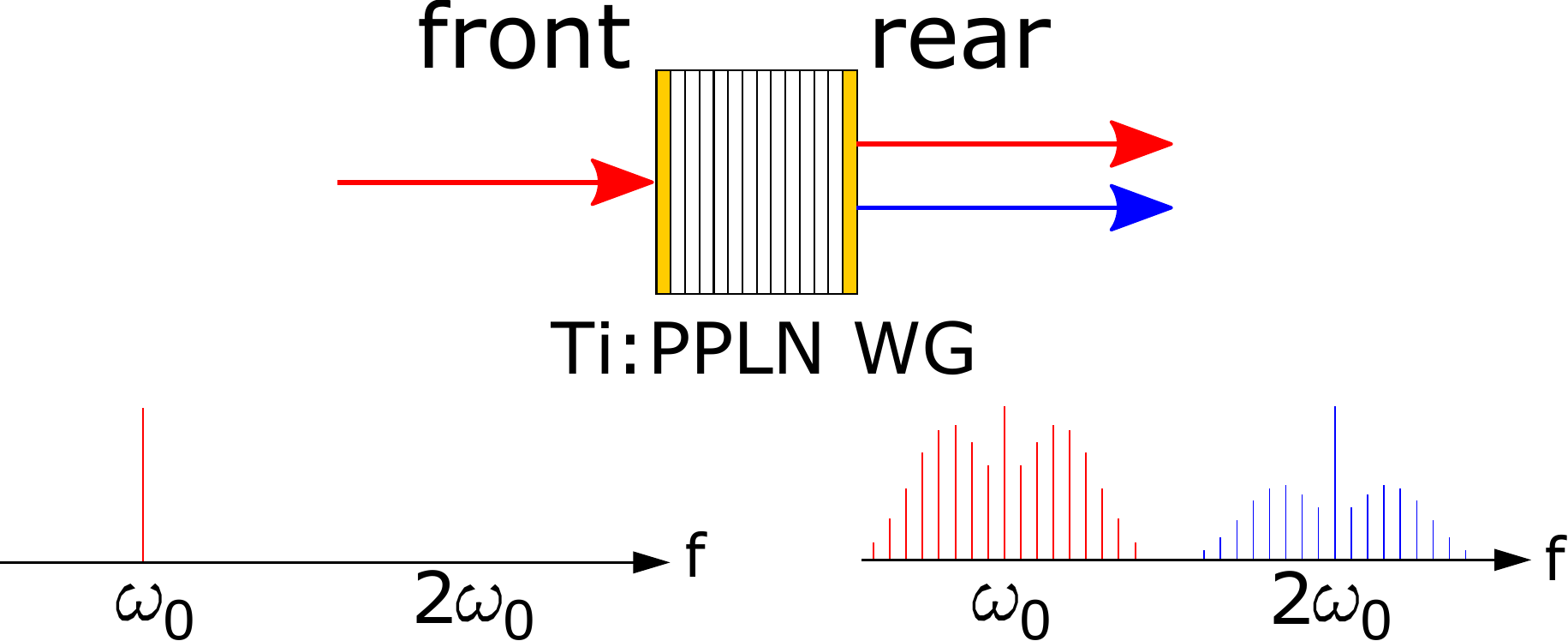}
		\caption{General scheme for CW pumped integrated OFC generation utilising the second-order nonlinearity. A titanium indiffused periodically poled lithium niobate waveguide (Ti:PPLN) with coatings deposited to the end faces, labeled front and rear, is pumped by a strong CW laser. Within the device an optical frequency comb (with a spectral spacing equal to the free spectral range of the system) is generated. Representative optical spectra for the input and output fields are shown.}
		\label{CombGen}
\end{figure}}

As with the more standard third-order systems, the natural progression is to move this technology to a miniaturized, integrated platform. In second-order systems, this has been previously investigated for single-pass pulsed driven systems \cite{Phillips11.OL,Langrock07.OL,Kowligy18.OL} and synchronously-pumped femtosecond OPO's \cite{McMahon16.S}. However, the possibility of generating optical frequency combs from second-order CW pumped integrated systems has only recently been experimentally investigated, the first observations being the work presented here \cite{Stefszky17Comb.A} and subsequently the published results of Ikuta \textit{et al} \cite{Ikuta18.OE}.

In this paper we present the first detailed experimental observation and extended theory of OFC generation from a CW-pumped, integrated second-order nonlinear system, paying particular attention to the limitations and anomalies that have been observed. A titanium-indiffused waveguide resonator is optically pumped at high powers in order to reach the regime where OFC generation is expected to occur. Under these conditions a comb envelope arising due to modulational instability (MI) is observed but is accompanied by self-pulsing and some form of mode competition. We compare the results to recent theory for comb generation in second-order nonlinear systems and show that operational conditions can be found that produce qualitatively similar structures.

\section{Theory}

\begin{figure}[!ht]
	\centering
	\includegraphics[width=15cm]{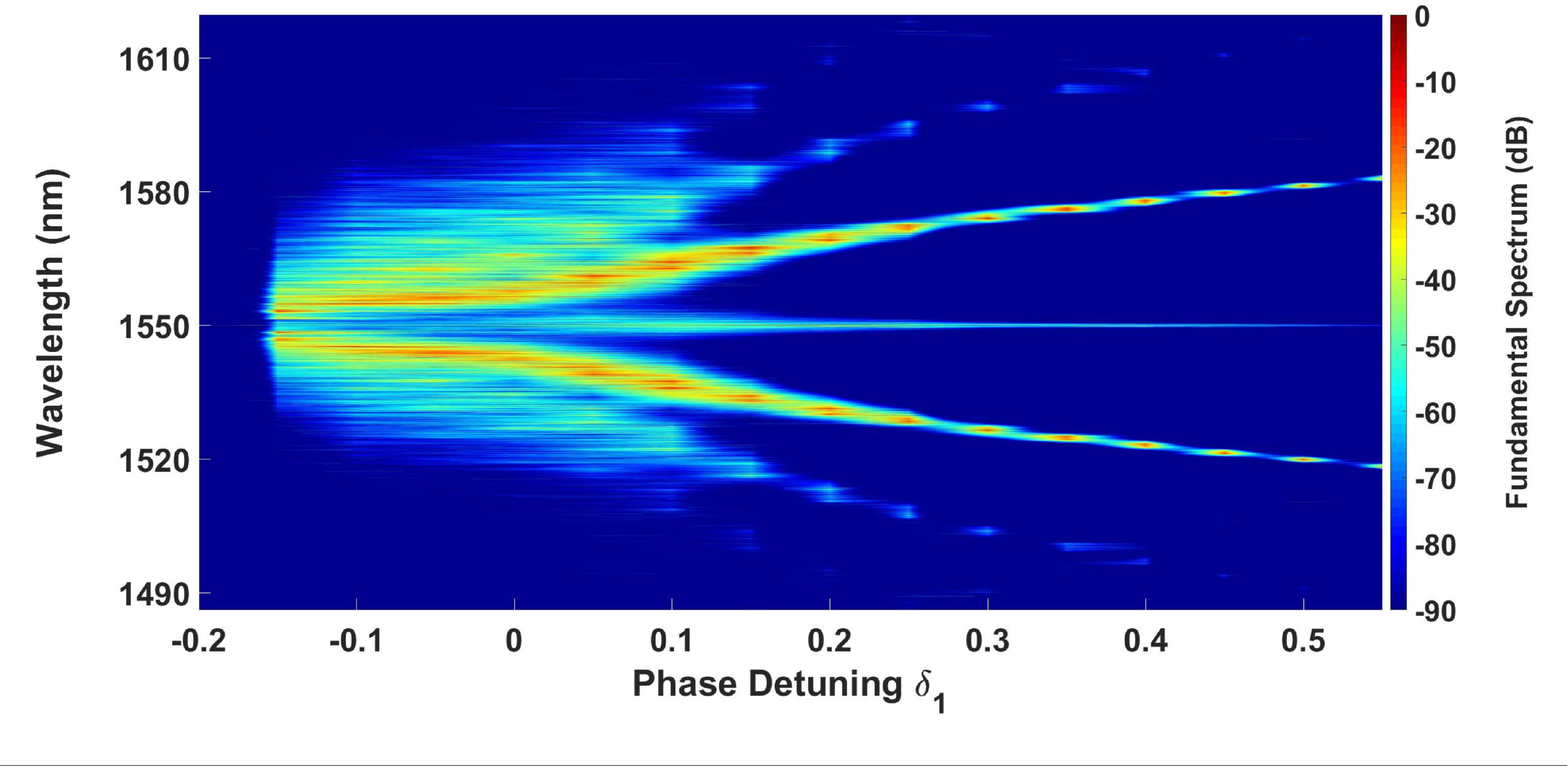}
	\caption{Spectral dynamics of waveguide sample as a function of the pump (fundamental) field detuning from cavity resonance with a 330mW pump field at 1550nm. The spectrum is normalized to the maximum power in the field for each detuning value.}
	\label{Dynamics}
\end{figure}

Much progress has recently been made towards understanding frequency comb generation from CW pumped second-order nonlinear materials. Leo {\it et al} presented a theory that includes temporal walk-off and showed that frequency combs can be generated in both singly resonant (in which only the pump field resonates)\cite{Leo16.PRL,Hansson17.PRA} and doubly resonant (in which both the pump and SH fields resonate)\cite{Leo16.PRA} systems. In fact, the dynamics of the system vary significantly for these two cases. This theory has been applied to the physical system from Ricciardi \textit{et al} \cite{Ricciardi15.PRA} and correctly predicts much of the observed behavior. However, the theory developed thus far works under the assumption that the intra-cavity second harmonic field is exactly zero after the mirror in which the pump field enters the system, thereby ensuring that the SH field is slaved to the fundamental field. This is not the case for our system, in which the pump field resonates and the second harmonic field exits the rear of the device (as shown in Fig. \ref{CombGen}) and it is therefore necessary to expand on this theory in order to model the system presented here. A full derivation of the equations and a description of all the working parameters, as well as the assumptions that are necessary for the modeling, can be found in the Appendix. The equations derived using this theory are solved using a split-step Fourier method and the spectral dynamics found through this expanded model when modeling our experimental device are presented in Fig. \ref{Dynamics}.

Fig. \ref{Dynamics} shows the predicted envelope of the comb structure (hence named the optical frequency comb envelope) produced by the waveguide resonator. The underlying comb structure, with a spacing equal to the free spectral range (8.3 GHz) of the cavity, is contained within this envelope, which arises due to modulational instability. This theory also allows for investigation of the temporal properties, although this is omitted here as we have no means of measuring signals that oscillate at several gigahertz \cite{Leo16.PRL}. The integrated approach presented here has a substantially higher nonlinear interaction strength (here $\kappa = 77 $ W$^{-1/2}$m$^{-1}$ as opposed to $\kappa \approx 11 $ W$^{-1/2}$m$^{-1}$ in the free space system \cite{Ricciardi15.PRA}).

\section{Experimental Setup}

The device is an 8mm long titanium-indiffused lithium niobate waveguide. The waveguide used for this work is made from indiffusion of a 7 $\mu$m titanium strip and is periodically poled using the electric-field periodic poling technique with a periodicity of 17.0 $\mu$m for phase matching at around 170 degrees Celsius. The full production details can be found in previous papers \cite{Stefszky17.PRA,Stefszky18.JO}. End-face coatings are applied to the sample in-house, resulting in a power reflectivity of 77$\pm$ 0.2 \% on the front face and 99.4$\pm$0.1 \% on the rear surface for the fundamental pump field. The propagation losses at the fundamental have been measured to be 0.16 $\pm$ 0.01 dB/cm. At the second harmonic wavelength, the front surface has a reflectivity over 99\% and is anti-reflection coated at the rear surface. The device is therefore double-pass for the second harmonic. 



\textbf{\begin{figure}[!ht]
		\centering
		\includegraphics[width = 10cm]{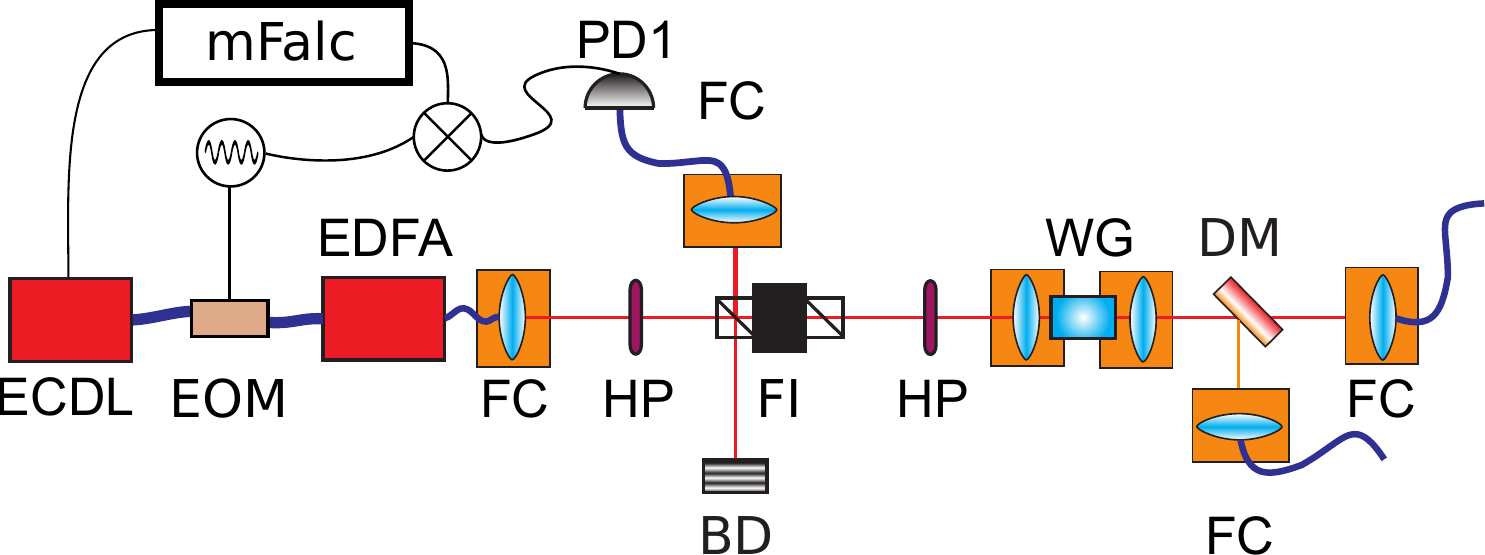}
		\caption{Experimental layout for the optical frequency comb generation measurements. ECDL denotes the Newport Velocity tunable (1520nm - 1570nm) ECDL laser source; EOM, electro-optic phase modulator; EDFA, Erbium-doped fiber amplifier; FC, fiber coupler; HP, Half-wave plate; BD, beam dump; FI, Faraday isolator; WG, waveguide sample; DM, dichroic mirror; PD, photodetector; mFalc, servo electronics for laser locking.}
		\label{FreqCombLay}
\end{figure}}

The experimental layout for driving the waveguide device and measuring the output optical powers and optical envelope spectra is shown in Fig. \ref{FreqCombLay}. A Newport Velocity tunable external-cavity diode laser (ECDL) is first fiber-coupled and directed through a fiber-coupled electro-optic phase modulator, driven at around 302MHz. The field exiting the modulator is then amplified in a home-built erbium doped fiber amplifier, providing up to 1 Watt of output optical power at around 1550nm. The laser field then exits the fiber and the power can be fine-tuned using a half-wave plate and Faraday isolator combination. The pump field is coupled to the waveguide mode with an overlap of approximately 86\%. The field that propagates backwards from the waveguide reflects at the Faraday isolator and is then fiber-coupled and detected with a fiber integrated photodiode (Thorlabs DET01CFC). This signal is mixed with the frequency modulation signal and low-pass filtered to produce the standard Pound-Drever-Hall error signal \cite{Black01.AJP} which is then shaped in the servo (Toptica mFalc) and fed back to the piezo of the ECDL cavity mirror. At the rear of the waveguide device both the fundamental field that leaks through the HR coating and the generated second-harmonic field are separated on a dichroic mirror. The optical power in these fields can be measured by placing a power meter after the dichroic mirror, or the fields can be fiber coupled and the optical envelope spectra measured using an interferometer based optical spectrum analyzer (OSA), model EXFO WA-1500 + WA650. It was necessary to use such a Fourier-transform based OSA due to the observed self-pulsing, as the Fourier-transform based analyzer allows for the measurement of all spectral components in a single shot, something that is not possible with the more standard scanning grating OSA.

\subsection{SHG Clamping}

Initial characterisation of the device is undertaken by measuring the optical SH power  produced on phasematching as the optical pump power is increased. In particular we search for the point at which SHG clamping is seen, the point at which further increases in the optical pump power no longer produce a corresponding gain in the generated SH power. Previous experiments and theory have both shown that this clamping is linked to the onset of OFC generation \cite{Ricciardi15.PRA, Hansson17.PRA}. The theoretical average power in $B_{out}$ is found when solving the full cavity map described in the Appendix. 


The pump power entering the waveguide is varied by rotating the half-wave plate in the beam path and/or varying the EDFA gain. The highest average SH power generated is found more easily by scanning the laser frequency over a complete cavity resonance (8.3 GHz scan at approximately 100Hz), thereby probing all possible cavity detunings, removing the need to stabilize the cavity. A power meter is then placed between the dichroic mirror and the fiber coupling (see Fig. \ref{FreqCombLay}), where the generated SH power as a function of cavity resonance condition is measured. The peak SH power generated in this way is then recorded and the results are show in Fig. \ref{SHGRun}. Waveguide coupling losses and optical losses due to the outcoupling lens and dichroic mirror are removed from these results.


\textbf{\begin{figure}[!ht]
		\centering
		\includegraphics[width = 8cm]{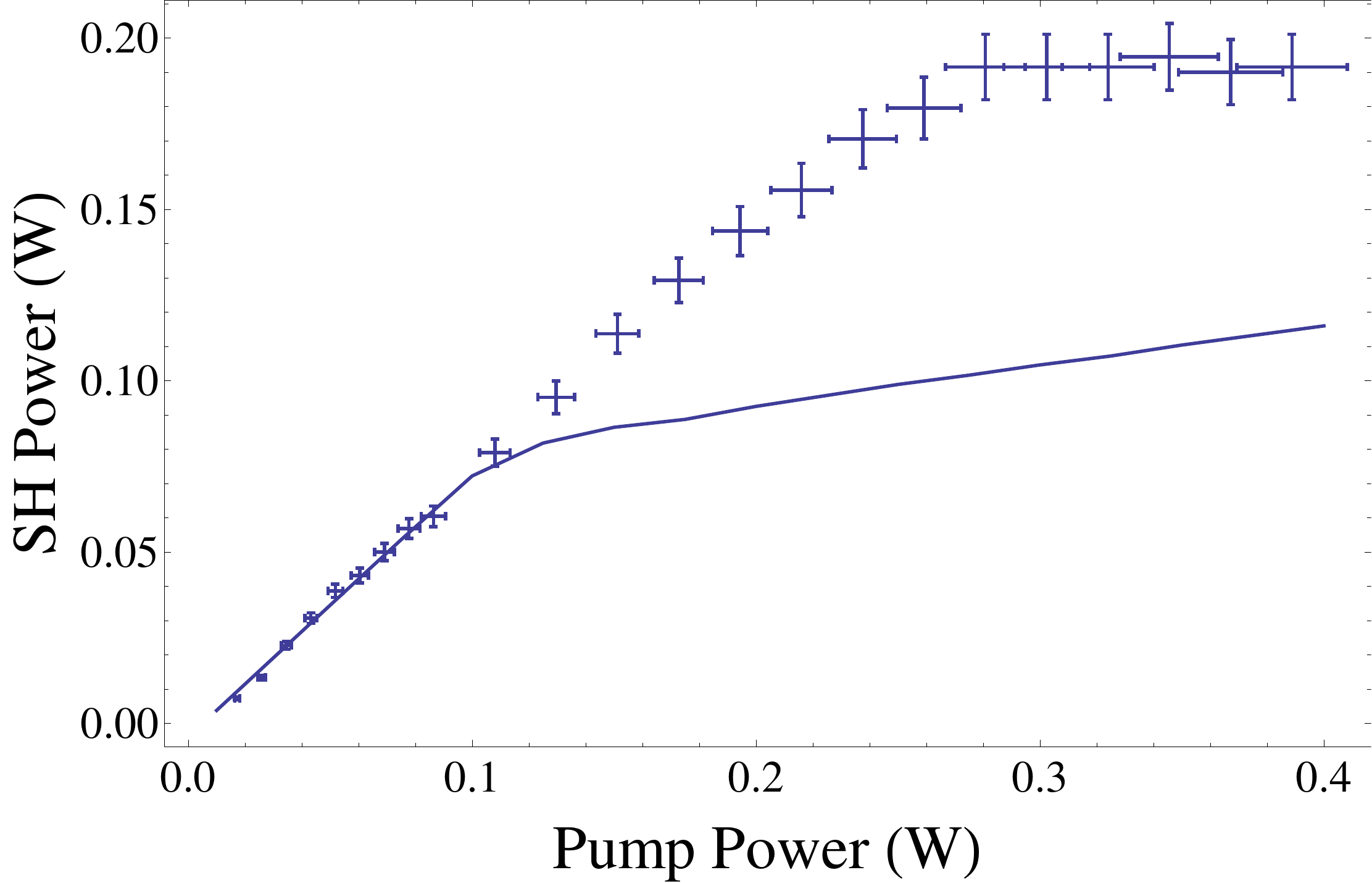}
		\caption{Example of a single SH power measurement run. The SH power exiting the rear of the waveguide device is recorded as the pump field power is varied. The trace shows the theoretically predicted means SH power with the nonlinear coupling coefficient fitted to the measured SH powers below an input pump power of 100mW.}
		\label{SHGRun}
\end{figure}}

A value for the nonlinear coefficient $\kappa$ is then found by fitting the theory to the data in which less than 100mW of pump power were used under the assumption that the laser wavelength is perfectly phase matched and that the cavity is perfectly on resonance. This results in a $\kappa$ value of 77 W$^{-1/2}$m$^{-1}$. 

There are two important features revealed in Fig. \ref{SHGRun} that should be noted. Firstly, it can be seen that the theory does not coincide with the measured SH powers at pump field powers greater than 100mW. The reason for this discrepancy is likely that one or more assumptions have broken down at the higher pump powers required to reach the expected clamping regime (the assumptions are detailed in the Appendix). For example, photothermal heating (the magnitude of which will depend on both the pump power and the cavity scan rate) will shift both the phasematching condition of the sample and the resonance condition of the cavity in a nontrivial way \cite{Carrascosa93.JAP}. It should also be mentioned that clamping was in fact occasionally seen in the region between 100mW and 300mW on separate measurements. This is consistent with the notion that reaching the clamping threshold for pump powers slightly above the expected pump power requires all system parameters to be near to their ideal values. This issue is further complicated by the presence of an observed mode competition, making it yet more difficult to reach the necessary conditions for reaching clamping threshold. The observed mode competition is described in detail in the following section.

The second interesting feature comes from the theory. It is seen that rather than having a strong clamping point at which the SH power no longer increases, as predicted by previous theoretical work for \cite{Leo16.PRA,Leo16.PRL,Hansson17.PRA}, instead the SH power growth is severely inhibited at the point where comb formation is expected to occur. The main difference between previous theoretical work and that presented here is that the second harmonic field is not zero at the point where the pump enters the system. Interestingly,  this SH growth inhibition has recently been experimentally observed in a second-order waveguide resonator experiment, even though the second harmonic power is zero at the point where the pump field enters \cite{Ikuta18.OE}.

\section{Device Stability}

Due to the necessarily high pump powers used and the demanding applications in mind, stability (both short and long term) is a key issue in frequency comb generation. Instabilities most often arise due to two competing nonlinear mechanisms with different signs, or from the interaction between two excited modes. In third order systems, for example, Xue {\it et al} have seen oscillations in a micro-ring resonator when scanning the resonance condition, that were believed to be due to an interaction between the pump and comb modes with a resonant mode at the second-harmonic frequency \cite{Xue17.LSA}. Much effort has been made towards gaining a further understanding of the processes behind such instabilities both in microresonators \cite{Liu14.O,Johnson06.OE} and free-space geometries \cite{Douillet99.JOSAB} as these issues must be overcome in order to produce a commercially viable system.

It should also be noted that in addition to the standard thermal instabilities that arise in both microresonator and waveguide architectures, the photorefraction seen in titanium indiffused waveguides is expected to further complicate the issue \cite{Becker85.APL}. To reduce the effect of photorefraction, the waveguide device is heated to around 180 degrees Celsius \cite{Carrascosa93.JAP}, but it is unknown exactly how much photorefraction remains. The limitations imposed by photorefraction have been investigated for single-pass, room temperature magnesium-oxide doped systems \cite{Li13.OE}, but it remains to be seen how the coexistence of both photorefraction and thermal instabilities affect device stability in a high temperature, undoped waveguide resonator.  In fact, it is expected that the amount of photorefraction will vary from waveguide to waveguide in a single sample due to the fact that the amount of photorefraction is dependent upon the number of impurities in the waveguide under question \cite{Carrascosa93.JAP}.

\subsection{Locked Cavity}

\begin{figure}[!ht]
	\centering
	\includegraphics[width = 8cm]{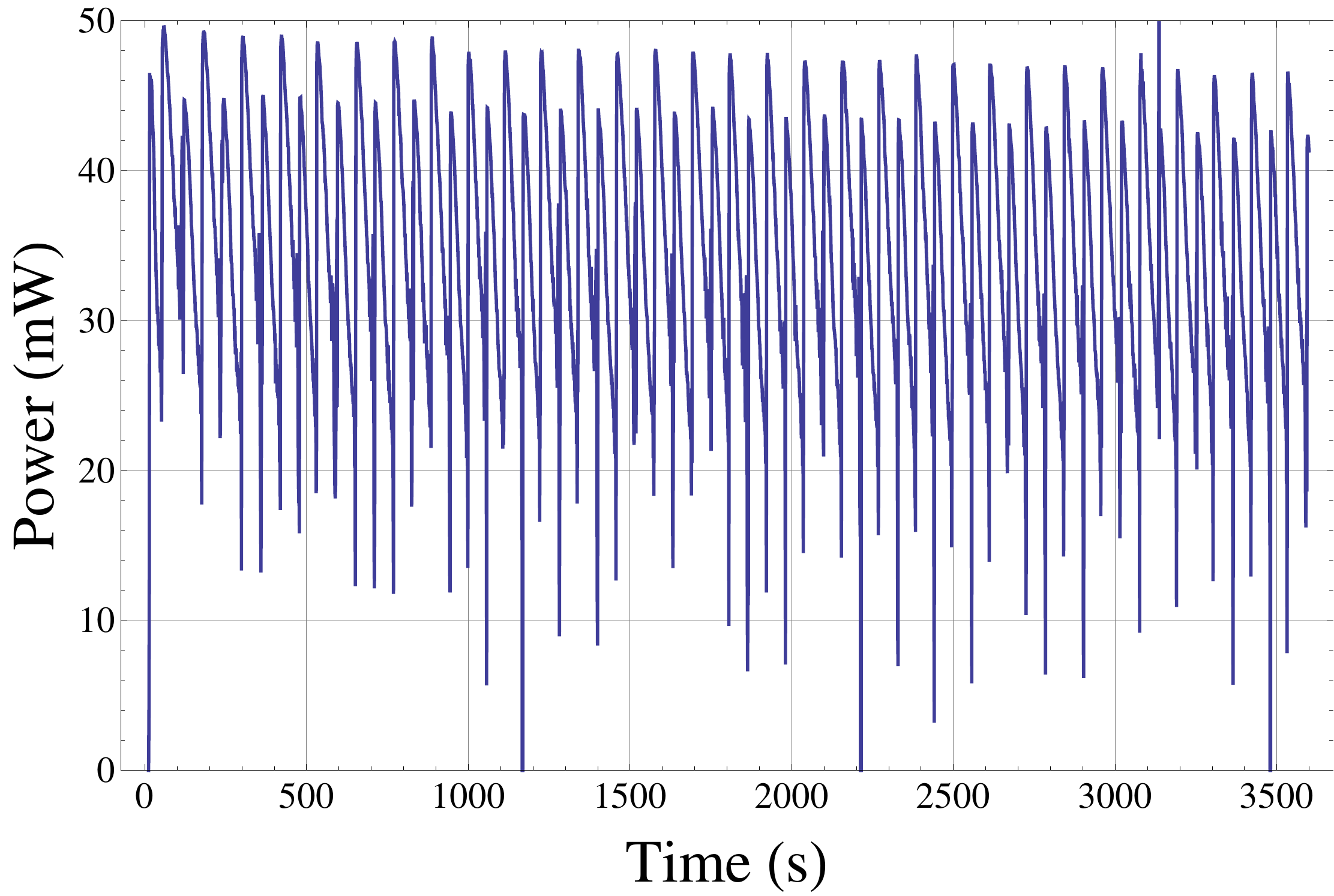}
	\caption{Measured SH power exiting the rear of the waveguide when the PDH lock is engaged. Coupled fundamental field power is approximately 100mW and the ECDL wavelength is set to 1550.2nm}
	\label{Lock120}
\end{figure}

We begin by investigating the behavior of the system when the PDH lock is engaged. The frequency of the pump laser is locked to the cavity length using the PDH scheme illustrated in Fig. \ref{FreqCombLay}. Below pump field powers of around 50mW the lock is very well behaved and is stable over hours \cite{Stefszky18.JO}. In fact, only minor drifts in the output SH power are observed, indicating that both heating and photorefraction play no major role at these powers. Were either of these effects present, one would expect the output SH power to drift because the error signal is derived only from the cavity length and not the phase matching condition. However, at higher pump powers, instabilities arise. Fig. \ref{Lock120} shows the measured SH power exiting the rear of the waveguide over one hour when the pump field power was set to approximately 100mW.

It is seen that the output SH power fluctuates significantly over time. Upon engaging the lock the measured SH output power jumps to a value around 45mW and then begins to decrease. After the power has reduced over some tens of seconds the power rapidly increases to a new maximum power level around 50mW, and once again begins to reduce. This process repeats over an hour, and in fact the frequency and maximum value of these jumps are quite consistent (as shown in Fig. \ref{Lock120}). Similar behavior is seen in the transmitted fundamental pump field. For further investigation, the output SH field was imaged onto a camera. It is seen that the spatial mode of this field drifts between two modes, in correlation with the observed oscillations in SH power. Monitoring the PDH error signal concurrently shows that the lock operates, for the most part, at the expected zero point. However, quick spikes in the signal are seen in correlation with the sharp jumps from low SH output powers to the local maxima in SH output power.

\subsection{Scanning over resonance}

\begin{figure}[!ht]
	\centering
	\includegraphics[width = 13cm]{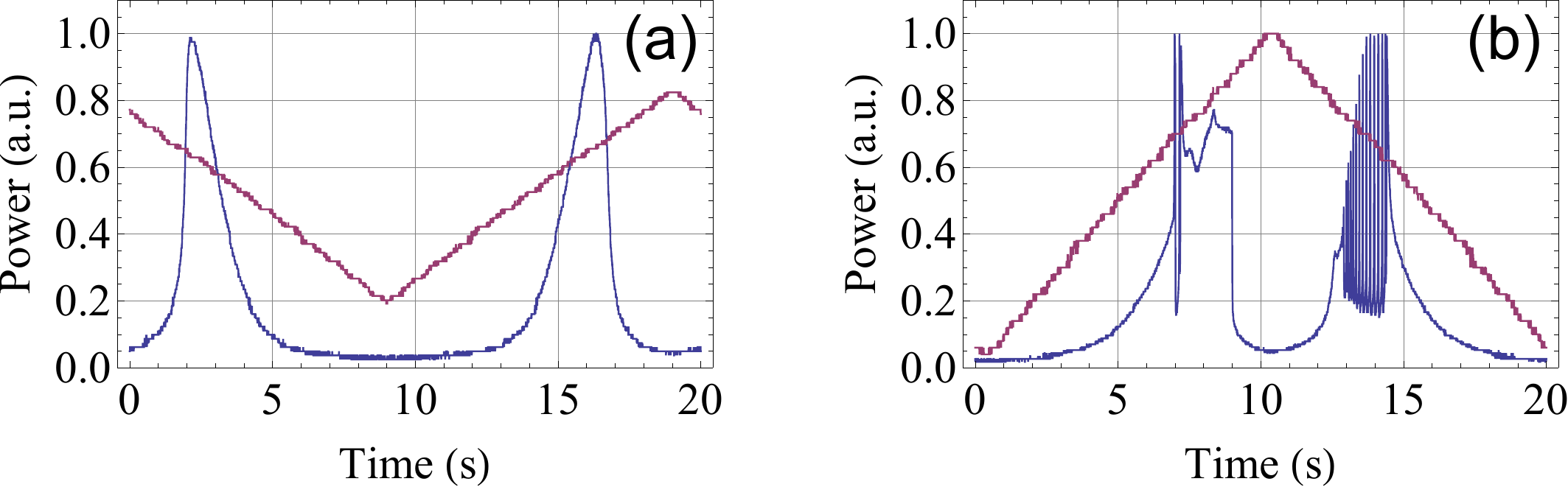}
	\caption{Transmitted optical pump field power measured as the wavelength of the pump  is scanned over the cavity resonance for 40mW (a) and 150mW (b) input pump powers. The scan voltage ramp is shown in purple, where a larger scan voltage corresponds to a red shift of the pump field. The pump field wavelength is scanned with a 50mHz ramp with a peak to peak amplitude of 0.3nm. SH powers are normalized to the maximum output power for each trace.}
	\label{CavScanPow}
\end{figure}

Another method for investigating the stability of the system is to scan over the resonance condition of the cavity. The wavelength of the pump laser is slowly scanned (approximately 0.03nm over 10s) over a single resonance while the transmitted power at around 1549.8 nm, the cold cavity phasematching wavelength, is monitored on a photodiode after the dichroic mirror. The transmitted power measured in this way is shown in Fig. \ref{CavScanPow} for two different pump powers.

We firstly note that the system behaves as expected when the pump field power entering the cavity is 40mW.  More specifically, the transmitted power is seen to be asymmetric around the resonance frequency due to heating of the cavity from the high circulating powers. However, once the incident pump field power exceeds around 150mW the transmitted pump field is seen to self-oscillate strongly over the scan. This behavior is most prominent when the laser frequency shifts from red to blue, as shown in Fig \ref{CavScanPow} (b).

Whilst a more in depth study of these results is left to a future paper, it is worth briefly mentioning that the results presented in this section seem to be consistent with a mode interaction between the desired SHG process and a second mode at the second harmonic frequency. The reasons for this hypothesis are threefold: firstly, detuning of the pump with respect to the cavity resonance is seen to unambiguously control the observed oscillations, secondly, the self-pulsing can be observed on both the fundamental and the second harmonic fields, and finally it is found that the output SH mode oscillates between two different spatial modes. Similar behavior has been attributed to thermal lensing effects in free-space systems \cite{Douillet99.JOSAB}. Recent work, however, has shown that even in magnesium oxide doped lithium niobate, where photorefraction is expected to be orders of magnitude weaker than in undoped lithium niobate, photorefraction is a much stronger effect than thermally induced nonuniformities \cite{Li13.OE}. The relative magnitudes of thermal effect and photorefraction at these elevated temperatures in the waveguides presented here remains to be seen.

\section{Optical Frequency Comb Generation}

In this section we present observations of optical comb envelope generation from a CW pumped \textit{integrated} device that utilizes the second-order nonlinearity. The integrated system provides a miniaturized platform that is mechanically robust, and substantially improves the field confinement, resulting in a much higher nonlinear interaction strength. In principal, this increase in nonlinear interaction strength reduces the amount of intra-cavity power required to reach the comb generation threshold.

\begin{figure}[!ht]
	\centering
	\includegraphics[width = 13cm]{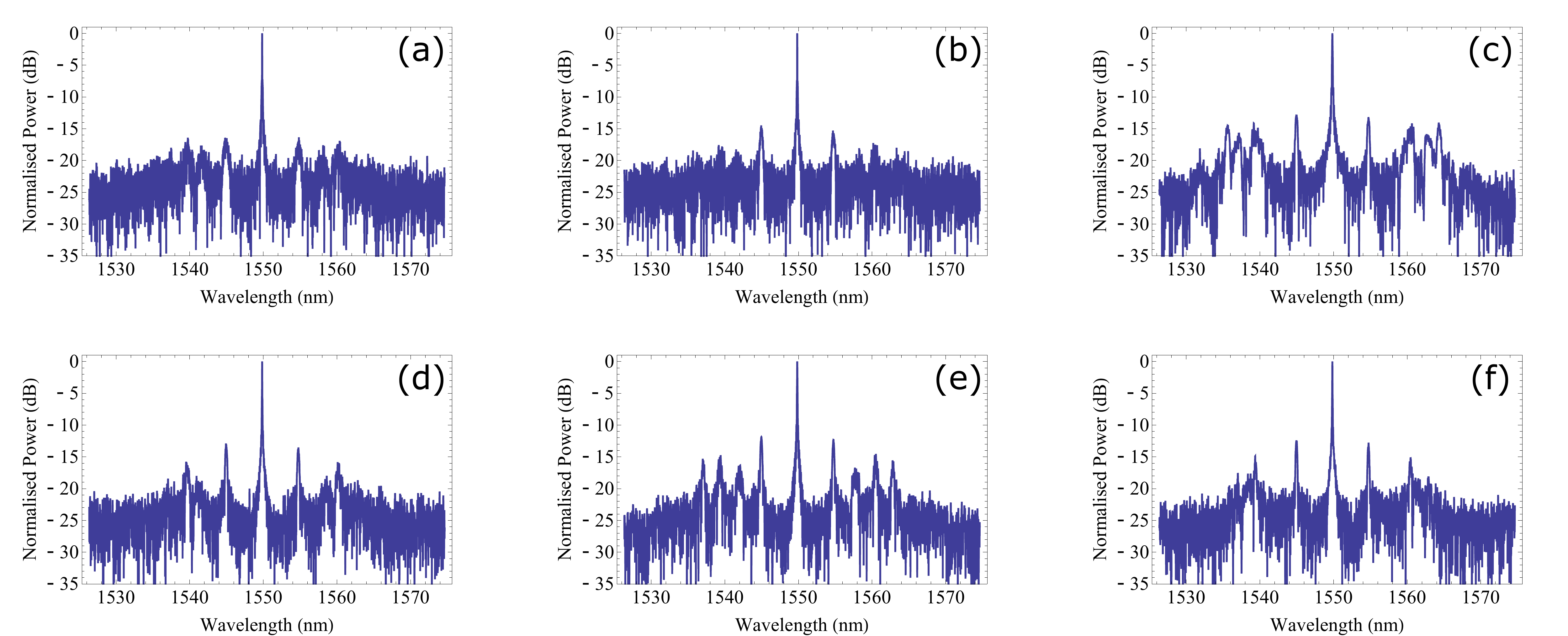}
	\caption{Measured optical spectra for a 330mW pump field power as the pump laser wavelength is scanned from red to blue. The traces are in chronological order, such that f) is measured when the pump laser is the most blue-shifted with respect to the cavity resonance. The pump field wavelength is 1549.8437nm.}
	\label{Spectra}
\end{figure}

In order to measure optical comb envelope generation around both the pump and SH fields, the pump power was set to 330mW and the pump wavelength was scanned across the cavity resonance, from red to blue, with a 100uHz ramp with an amplitude of 0.3nm. Optical comb envelope generation is seen when pumping with powers of around 200mW and above. Whilst this value is below the clamping demonstrated in Figure \ref{SHGRun}, it is important to again note that clamping was in fact occasionally seen in the region between 100mW and 300mW. The spectrum of the fundamental field exiting the rear of the waveguide, after fiber coupling, is measured in the Fourier-transform-spectrometry based optical spectrum analyzer. A selection of measured optical spectra for different cavity detunings $\delta_1$, are shown in Fig. \ref{Spectra}. 

\begin{figure}[!ht]
	\centering
	\includegraphics[trim={0 7cm 0 7cm},width = 14cm]{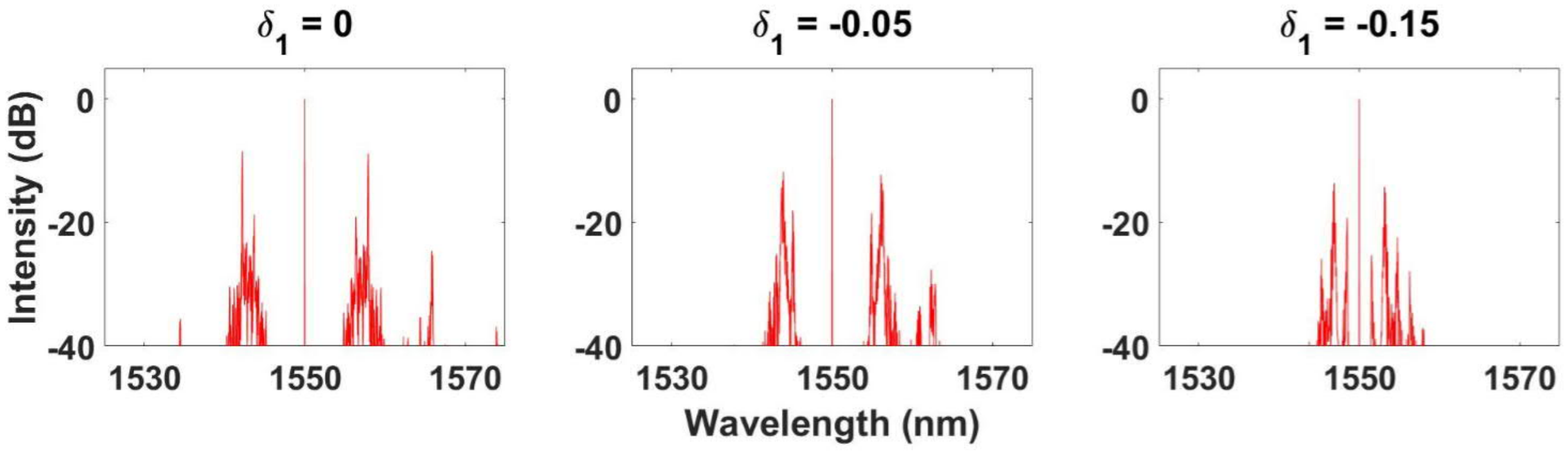}
	\caption{Predicted pump field spectra with a pump field power of 330mW and decreasing cavity detunings (corresponding to a laser frequency scan from red to blue) of $\delta$ = 0 (a), -0.05 (b) and -0.15 (c).}
	\label{TheorySpectra}
\end{figure}

The measured optical envelope spectra shown in Fig. \ref{Spectra} are taken from a single sweep over the cavity resonance as the pump laser wavelength is scanned (from red to blue) over the cavity resonance. As the pump laser wavelength scan approaches the cavity resonance condition, the optical comb envelope generation is seen to occur more often, and the generated features are stronger and have more structure. At the edge of resonance the structure that is typically seen is a broadening of the central wavelength and 2 small frequency comb envelopes (which we henceforth name MI peaks) equidistant at around 3nm from the pump wavelength. These peaks reach a maximum magnitude of about -10dBc. To aid in comparison between these results and the expected theory, a number of spectra for a selection of cavity detunings $\delta_1$ in Fig. \ref{Dynamics} are presented in Fig. \ref{TheorySpectra}. 

Qualitatively, the measured traces show some similar behavior to the theoretical predictions presented in Fig. \ref{TheorySpectra}. The theory predicts that main MI peaks should form between 2nm to 7nm from the carrier wavelength and this is seen in the measured spectra.

However, some features in the pump field optical spectra that do not coincide with the theory are also observed. A number of the measured optical spectra show more structure than is predicted, particularly when the pump wavelength is close to cavity resonance (Fig. \ref{Spectra} c-f). Furthermore, the spectral peak at the pump wavelength is seen to broaden for the majority of spectra, again with the effect strongest when the pump field is closest to resonance. Whilst this behavior is present in the singly resonant theory for certain detunings, as seen in Fig. \ref{Dynamics}, the expected magnitude of this effect is much less than is observed here. 

It is known that non-stationary inputs to Fourier-transform spectrometers can create artificial broadening around the main spectral peak and hence this effect has been investigated. In order to investigate the influence of this effect on our measurement, we chopped the laser power before it enters the optical spectrum analyzer such that the response of the device could be investigated. It was seen that an oscillating input power primarily generates sidebands that are spaced from the central laser frequency by $\approx$4.5 nm per kilohertz. The fastest oscillations in the fundamental power seen when the system approached resonance and began to oscillate (in a similar way to that shown in Fig. \ref{CavScanPow}) were not more than around 100Hz. Therefore, we expect to see broadening up to around 0.5nm due to the non-stationary nature of the input signal. This is consistent with the observed broadening of the central frequency, for example in Fig. \ref{Spectra} (c), and we therefore conclude that \textit{the observed broadening of the central peak is almost certainly a measurement artifact}.

The observed MI peaks cannot be attributed to such a measurement effect. To further ensure that the measured MI structure seen in Fig. \ref{Spectra} were not merely measurement artifacts the spectra were also measured in a standard scanning grating optical spectrum analyzer. Similar features to those presented in Fig. \ref{Spectra} were observed, albeit only over part of the scan due to the relatively long measurement times required, providing confidence that the observed MI peaks are not a measurement artifact. We also note that one possible explanation for the more complicated spectra is that the observed oscillations in output power of the system may lead to a time-dependent cavity detuning. It is therefore possible that a single measurement may integrate over some range of cavity detunings, thereby leading to more structure than one might expect from a single value for the cavity detuning. This picture is further complicated by the above-mentioned self pulsing effect that is not accounted for by the present theoretical model.

\begin{figure}[!ht]
	\centering
	\includegraphics[width = 13cm]{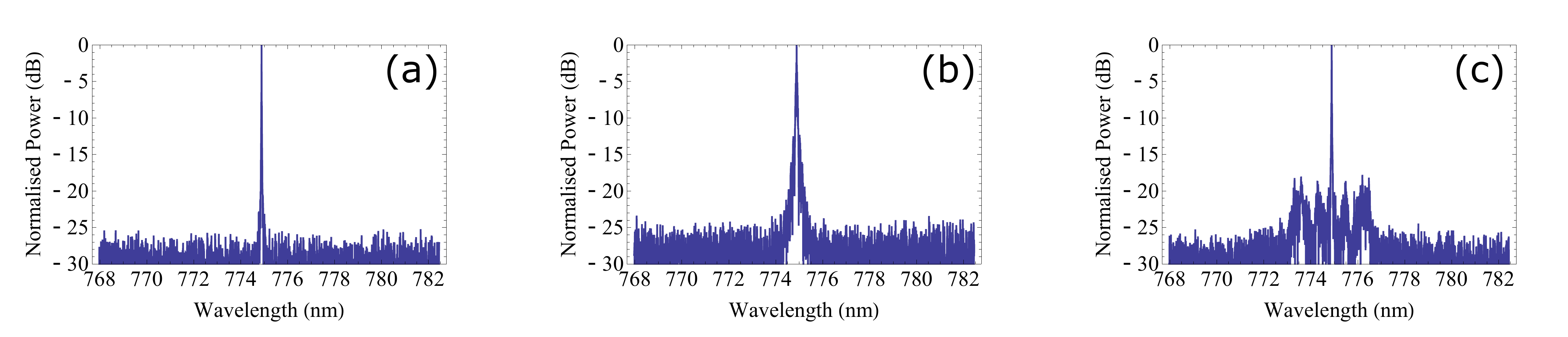}
	\caption{Spectra of the SH field with a 330mW pump field at 1549.78nm. The shown spectra show sample spectra when the system is far from resonance (a), near to resonance and producing large amounts of SH power (b) and very near to resonance (c).}
	\label{SHGSpectra}
\end{figure}

Next, the optical spectrum of the fiber coupled SH field is also measured using the interferometer based optical spectrum analyzer. Similarly to the measurements for the fundamental (pump) field, the wavelength of the pump field is scanned over resonance, from red to blue, with a 100uHz ramp with an amplitude of 0.3nm. A selection of measured SH spectra are shown in Fig. \ref{SHGSpectra}.

The SH field spectra shown in Fig. \ref{SHGSpectra} illustrate the three types of behavior that were observed: a) shows a reference spectrum, b) shows broadening of the central laser frequency of the SH field, and c) shows a rarely seen structure that occurs only when the pump field wavelength is very near to resonance. The observed broadening of the central peak is again accompanied by oscillations in the SH power and are thus likely an artifact of the measurement. Theory predicts the generation of MI peaks in the spectrum appearing equidistant from the main SH field signal, with a magnitude of approximately -30dBc. Note that this is below the noise floor of our measurements, and additionally have the same structure as shown for the pump field (see Fig. \ref{TheorySpectra}) and as such these figures are not shown. Similarly to the fundamental field spectra, Fig. \ref{SHGSpectra} (c) shows spectral structure in the SH field that is not predicted by the theory.

\section{Discussion and Conclusion}

We have presented a CW pumped \textit{integrated} second-order nonlinear waveguide cavity device that exhibits the optical comb envelope generation. We have shown that, in agreement with the extended theory of Leo \textit{et al} \cite{Leo16.PRL} that is developed in this paper, that MI peaks in the fundamental field spectrum are generated around the central laser frequency. Interestingly, the optical spectra of both the pump and SH fields near to cavity resonance often show more structure than is predicted by the theory. In an attempt to stabilize the system, a Pound-Drever-Hall locking scheme was constructed to lock the pump laser frequency to the cavity resonance condition. Whilst this scheme worked well at low powers, the system becomes unstable at higher pump powers and the transmitted fields are seen to oscillate, similar to previous results in free-space systems \cite{Ricciardi15.PRA}.

These results indicate that the CW pumped titanium indiffused waveguide resonator is a promising platform for further investigations into OFC generation. This platform offers a high field interaction strength and low losses, leading to low threshold powers, and also offers the possibility of tailoring the resonance condition of both the fundamental and SH fields. However, further investigation of the device requires a more complete understanding of the mechanisms that produce the observed instabilities; mode-coupling, photothermal effects and photorefractive effects. It is evident that a more complete theory including photorefraction and/or photothermal effects may be necessary to full describe some of the more complicated spectral features observed from this system.

\begin{acknowledgments}
 The authors are grateful to the University of Helsinki, the Academy of Finland, Tekes - the Finnish Funding Agency for Innovation, the Emil Aaltonen Foundation and the DFG (Deutsche Forschungsgemeinschaft) via the Gottfried Wilhelm Leibniz-Preis for funding this research. We thank Yauhen Baravets for help in building the EDFA amplifier and Mikko Lotti for providing the split-step Fourier method software.
\end{acknowledgments}

\section{Appendix A: Full Cavity Map}

Here we detail the theoretical analysis of the waveguide resonator system. Following the same method as that use be Leo \textit{et al} \cite{Leo16.PRA} the fundamental and second harmonic intracavity fields, $A_{m+1}(0,\tau)$ and $B_{m+1}(0,\tau)$, at the beginning ($z=0$) of the $(m+1)$th roundtrip can be related to the fields at the end ($z=L$) of the $m$th round-trip, $A_{m}(L,\tau)$ and $B_{m}(L,\tau)$. However, the situation is slightly more complicated in the presented device due to the fact that the two fields couple out of the cavity at separate mirrors. The physical system has been shown in Fig. \ref{CombGen} while the equivalent theoretical system is illustrated in Figure \ref{CavSchem}.

\textbf{\begin{figure}[!ht]
		\centering
		\includegraphics[width = 6cm]{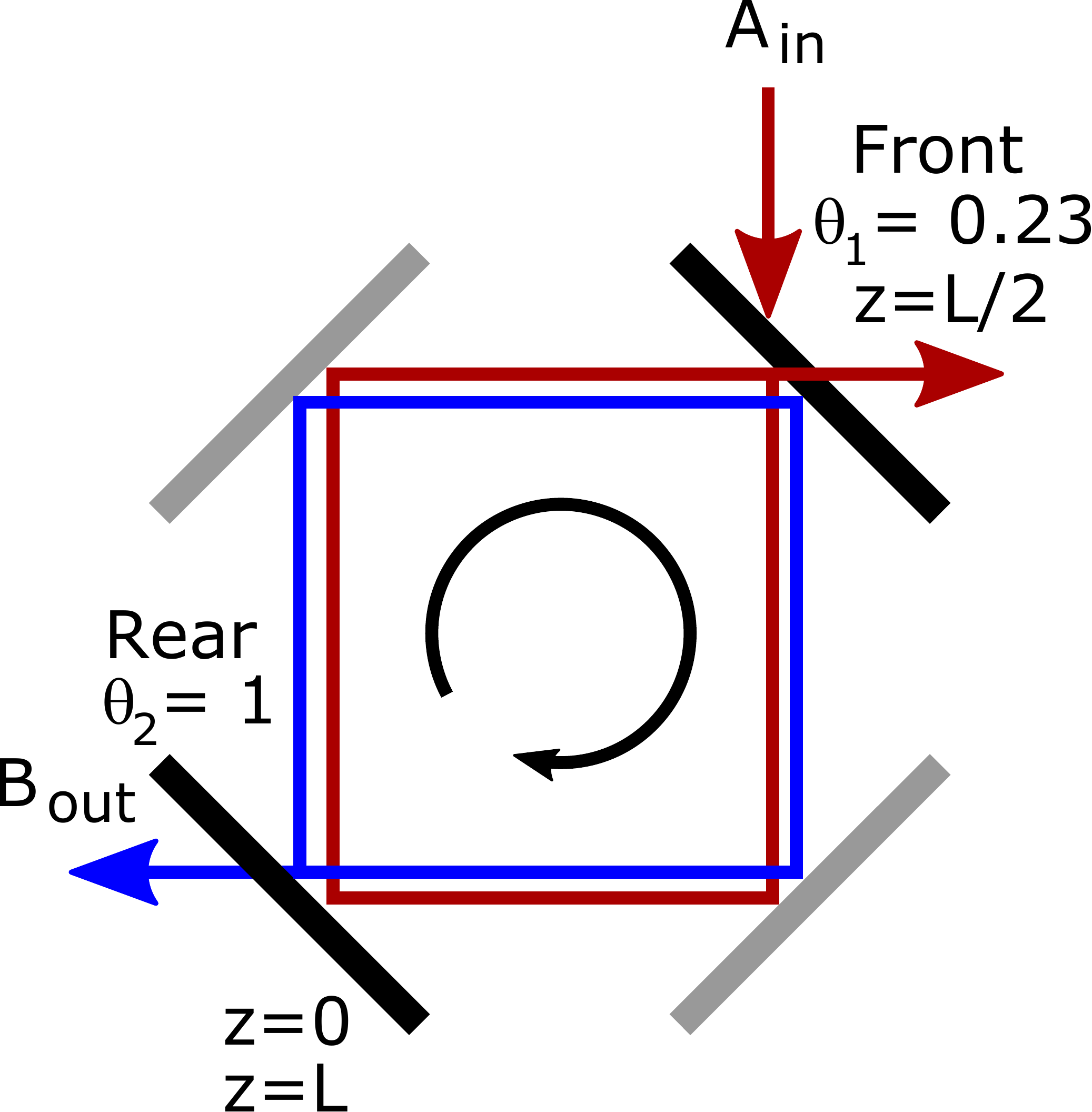}
		\caption{Equivalent cavity for the monolithic waveguide resonator. The cavity is unfolded using two perfect mirrors (in grey) such that the dynamics from z=0 to z=L/2 and the dynamics from z=L/2 to z=L can be considered. Note that system described in such a way is now a traveling-wave geometry.}
		\label{CavSchem}
\end{figure}}

It is therefore necessary to separate the evolution of the fields into two steps, describing the evolution from the rear mirror to the front mirror and back again. Choosing $z=0$ to correspond to the rear mirror of the device, one can write 
\begin{eqnarray}
A_{m,aux}(L/2,\tau) &=& \sqrt{1-\theta_1}A_m(0\rightarrow L/2, \tau)e^{-i \delta_1/2}+\sqrt{\theta_1}A_{in} \\
A_{m+1}(0,\tau) &=& A_{m,aux}(L/2\rightarrow L,\tau)e^{-i \delta_1/2} \\
B_{m,aux}(L/2,\tau) &=& \sqrt{1-\theta_2}B_{m}(0 \rightarrow L/2,\tau)e^{-i \delta_2/2}\\
B_{m+1}(0,\tau) &=& B_{m,aux}(L/2 \rightarrow L,\tau)e^{-i \delta_2/2},
\end{eqnarray}
where $\theta_1=0.23$ and $\theta_2=1$ are the power transmittivity for the fundamental and second harmonic fields on the front and rear mirrors, respectively, $\delta_1$ and $\delta_2$ are the phase detunings of the intracavity fundamental and second harmonic fields, respectively, from the nearest cavity resonance, $A_{in}$ is the continuous-wave driving field at the fundamental frequency. The auxiliary fields $A_{m,aux}(L/2,\tau)$ and $B_{m,aux}(L/2,\tau)$ are used to describe propagation from the front mirror to the rear mirror of the fundamental and second harmonic fields respectively. Note that it is assumed that the rear mirror is a perfect reflector for the fundamental field and that it perfectly transmits the second harmonic field, while it is assumed that the front mirror is a perfect reflector for the second harmonic field, whilst transmitting 23\% of the fundamental.

The evolution of the fields described over a single pass of the waveguide resonator can then be described using \cite{Leo16.PRA}
\begin{eqnarray}
\frac{\partial A_{m,aux}}{\partial z} &=& \left[-\frac{\alpha_{c1}}{2}-i\frac{k_1''}{2}\frac{\partial^2}{\partial\tau^2}\right] A_{m,aux} +i \kappa B_m A^*_x e^{-i \Delta k z} \\
\frac{\partial A_m}{\partial z} &=& \left[-\frac{\alpha_{c1}}{2}-i\frac{k_1''}{2}\frac{\partial^2}{\partial\tau^2}\right] A_m +i \kappa B_m A^*_x e^{-i \Delta k z} \\
\frac{\partial B_{m,aux}}{\partial z} &=& \left[-\frac{\alpha_{c2}}{2}-\Delta k'\frac{\partial}{\partial\tau}-i\frac{k_2''}{2}\frac{\partial^2}{\partial\tau^2}\right] B_{m,aux} +i \kappa A^2_m e^{-i \Delta k z}\\
\frac{\partial B_m}{\partial z} &=& \left[-\frac{\alpha_{c2}}{2}-\Delta k'\frac{\partial}{\partial\tau}-i\frac{k_2''}{2}\frac{\partial^2}{\partial\tau^2}\right] B_m +i \kappa A^2_m e^{-i \Delta k z}.
\end{eqnarray}
This set of equations is solved using a split-step Fourier method in order to find both the resulting pulse waveforms and the spectral dynamics of the system. By choosing the correct parameters the results from Leo \textit{et al} have been faithfully reproduced by the theory presented here \cite{Leo16.PRL}, giving us confidence in the results obtained by this extended model.

The remaining parameters and their values are as follows: the pump laser wavelength is 1550 nm, the round-trip length is $L =16$ mm, the cavity finesse has been measured to be 20 \cite{Stefszky17.PRA} which gives a fundamental loss parameter of $\alpha_1 = \pi/20$, the group velocity dispersion (GVD) of bulk lithium niobate at the pump field wavelength is $k''_1 = 0.107$ ps$^2$ m$^{-1}$ and at the SH field wavelength is $k''_2 = 0.405$ ps$^2$ m$^{-1}$ \cite{Jundt97.OL}, the temporal walk-off is $\Delta k' = 318$ ps m$^{-1}$, the wave-vector mismatch is $\Delta k = 2k_1-k_2-2 \pi / \Lambda$, where $k_1$ and $k_2$ are the wave vectors of the fundamental and pump fields respectively and $\Lambda$ is the component due to the quasi-phase matching grating, the round-trip propagation loss for the fundamental field is $\alpha_{c1} = (2 \alpha_1-\theta_1)/L$ \cite{Stefszky18.JO}, the propagation loss for the second harmonic field is assumed to be double that for the fundamental $\alpha_{c2} = 2\alpha_{c1}$, and the nonlinear coupling coefficient is $\kappa = 77 $ W$^{-1/2}$m$^{-1}$.The spectral dynamics of the system are found for a pump field power of 330mW, equal to that used in the experiment.

\section{Appendix B: Modeling Assumptions}

Here we note that there are a number of additional assumptions that are made in order to simplify the theoretical treatment of the presented system. These assumptions are: that the nonlinear coefficient is real, that the standing-wave cavity is equivalent to a traveling-wave cavity, and that the phase detuning relationship between the fundamental and second harmonic fields is given by $\delta_2 \approx 2\delta_1$. Firstly, in order for the nonlinear coefficient to be real in the double-pass cavity design, there must be perfect constructive interference between the forward and reverse-pass waves. It can be shown that a double-pass device with a real nonlinear coefficient (or equivalently perfect constructive interference) produces the same amount of SHG as a single-pass device with twice the length \cite{Imeshev98.OL}. Previous measurements of the device presented here show that this phase relationship is in fact met, giving us confidence that this parameter can be treated as a real number, thereby simplifying the treatment \cite{Stefszky18.JO}.

Next, we argue that the standing-wave design presented here is equivalent to the traveling-wave design considered in the theory (Fig. \ref{CavSchem}) under two assumptions; that the forward and reverse waves do not interact, and that thermal effects and photorefraction can be neglected. Previous theoretical work has also neglected thermal effects. However, it is possible that thermal effects are more pronounced in the waveguide resonator due to higher absorption and the standing-wave geometry. Furthermore, the titanium indiffused lithium niobate resonator presented here is susceptible to photorefraction. Experimentally, these effects have been reduced as much as possible by heating the sample to around 170 degrees Celsius (thereby reducing photorefraction \cite{Becker85.APL}) and by choosing a low-loss waveguide. Inclusion of these effects in the theory would require significant effort and is left as future work. Interaction between the forward and reverse waves in the standing cavity does not occur because the phase matching condition is not fulfilled.

Finally, natural phase-matching is typically assumed in the systems that have been previously investigated \cite{Leo16.PRA,Leo16.PRL,Hansson17.PRA}. It is easy to show that this system allows one to relate the cavity phase detunings of the two fields via $\delta_2 = 2 \delta_1$ \cite{Leo16.PRA}. However, it is not obvious that this should also be the case for a quasi-phase-matched system, such as that presented here. The relationship can be found by calculating the phase shift per round trip of the two fields, $\phi = k L$, where $L$ is the cavity length and $k$ is the wavevector. This value can be well approximated in the waveguide system using the Sellmeier equations for bulk lithium niobate \cite{Jundt97.OL}. We find that $\delta_2 \approx 2.087 \delta_1$ for the entire pump wavelength range of interest. For simplicity, this relationship has been approximated to $\delta_2 \approx 2 \delta_1$ as the dynamics were not seen to vary significantly with minor changes in this relationship.


%

\end{document}